\documentstyle[amsmath,amssymb,12pt,aaspp4,flushrt]{article}

\newcommand{\hMpc}{\ifmmode{h^{-1}{\rm Mpc}}\else{$h^{-1}{\rm Mpc}$}\fi}
\newcommand{\cD}{{\mathcal{D}}}
\newcommand{\cF}{{\mathcal{F}}}
\newcommand{\cP}{{\mathcal{P}}}
\newcommand{\rd}{{\mathrm{d}}}
\makeatletter

\def\@biblabel#1{}
\def\@bcite#1#2{(#1\if@tempswa , #2\fi)}
\def\@pcite#1#2{#1\if@tempswa , #2\fi}
\def\@citefmta#1#2{#1 (#2)}
\def\@citefmtb#1#2{#1 #2}
\let\citefmt=\@citefmta
\def\@citex[#1]#2{\if@filesw\immediate\write\@auxout{\string\citation{#2}}\fi
  \def\@citea{}\@cite{\@for\@citeb:=#2\do
    {\@citea\def\@citea{;\penalty\@m\ }\@ifundefined
    {b@\@citeb}{{\bf ?}\@warning
{Citation `\@citeb' on page \thepage \space undefined}}%
{\csname b@\@citeb\endcsname}}}{#1}}

\def\cite{\@ifnextchar [{\let\citefmt=\@citefmtb
                          \let\@cite=\@bcite\@tempswatrue \@citex}
                        {\let\citefmt=\@citefmtb
                          \let\@cite=\@bcite\@tempswafalse \@citex[]}}
\def\pcite{\@ifnextchar [{\let\citefmt=\@citefmtb
                          \let\@cite=\@pcite\@tempswatrue\@citex}
                        {\let\citefmt=\@citefmtb
                          \let\@cite=\@pcite\@tempswafalse\@citex[]}}
\def\scite{\@ifnextchar [{\let\citefmt=\@citefmta
                          \let\@cite=\@pcite\@tempswatrue\@citex}
                        {\let\citefmt=\@citefmta
                          \let\@cite=\@pcite\@tempswafalse\@citex[]}}

\makeatother

\begin{document}

\title{ Disentangling the Cosmic Web I: \\ 
Morphology of Isodensity Contours }

\author{
\newcounter{fusspilz}\setcounter{fusspilz}{0}
\def\fusspilz{\stepcounter{fusspilz}\fnsymbol{fusspilz}}
Jens Schmalzing$^{1,2,\fusspilz}$,
Thomas Buchert$^{3,1,\fusspilz}$,
Adrian L.~Melott$^{4,\fusspilz}$,
Varun Sahni$^{5,\fusspilz}$,
B.~S.~Sathyaprakash$^{6,\fusspilz}$, and
Sergei~F.~Shandarin$^{4,7,\fusspilz}$
\renewcommand{\thefootnote}{\arabic{footnote}}
\footnotetext[1]{
Ludwig--Maximilians--Universit\"at,
Theresienstra{\ss}e 37,
80333 M\"unchen, Germany. }
\footnotetext[2]{
Max--Planck--Institut f\"ur Astrophysik,
Karl--Schwarzschild--Stra{\ss}e 1,
85740 Garching, Germany. }
\footnotetext[3]{
Theory Division,
CERN,
1211 Gen\`eve 23, Switzerland. }
\footnotetext[4]{
Department of Physics and Astronomy,
University of Kansas,
Lawrence, Kansas 66045, USA. }
\footnotetext[5]{
Inter--University Centre for Astronomy and Astrophysics,
Post Bag 4, Pune 411007, India. }
\footnotetext[6]{
Department of Physics and Astronomy,
University of Wales, College of Cardiff,
Cardiff CF2 3YB, UK. }
\footnotetext[7]{
Theoretical Astrophysics Center,
Juliane Maries Vej 30,
2100 K{\o}bnhavn \O, Denmark. }
\renewcommand{\thefootnote}{\fnsymbol{footnote}}
\footnotetext[1]{email jensen@mpa-garching.mpg.de }
\footnotetext[2]{email buchert@theorie.physik.uni-muenchen.de }
\footnotetext[3]{email melott@kusmos.phsx.ukans.edu }
\footnotetext[4]{email varun@iucaa.ernet.in }
\footnotetext[5]{email spxbss@astro.cf.ac.uk }
\footnotetext[6]{email sergei@kusmos.phsx.ukans.edu }
\renewcommand{\thefootnote}{\arabic{footnote}}
\setcounter{footnote}{5}
}

\begin{abstract}
We  apply  Minkowski  functionals  and  various  derived  measures  to
decipher the  morphological properties of  large--scale structure seen
in simulations  of gravitational evolution.   Minkowski functionals of
isodensity contours  serve as tools  to test global properties  of the
density field.   Furthermore, we identify coherent  objects at various
threshold  levels and calculate  their partial  Minkowski functionals.
We propose  a set of  two derived dimensionless  quantities, planarity
and  filamentarity, which  reduce the  morphological information  in a
simple and  intuitive way.   Several simulations of  the gravitational
evolution  of  initial  power--law  spectra provide  a  framework  for
systematic tests of our method.
\end{abstract}

\begin{keywords}
{methods: numerical; methods: statistical; dark matter; large-scale
structure of Universe }
\end{keywords}

\section{Introduction}

The existence  of the large--scale clustering of  galaxies had already
been well established by the early 1970's mainly due to the pioneering
work of {\scite{totsuji:correlation}} and {\scite{peebles:nature}} who
showed that  the two--point correlation  function for galaxies  in the
Lick and  Zwicky catalogues  was positive and  had the  power-law form
$\xi(r)\propto{r^{-1.8}}$  on scales $\lesssim10\hMpc$.   Their result
was later  extended to  three--dimensional galaxy catalogues  as well.
Although  the clustering  of galaxies  is  now a  well--known fact,  a
complete  description  of clustering  which  includes its  geometrical
features has  so far eluded researchers.   This is perhaps  due to the
fact  that the galaxy  density field  which we  observe appears  to be
strongly non--Gaussian.  A Gaussian random field is uniquely described
by its  power spectrum $P(k)$, or its  two--point correlation function
$\xi(r)$  since $\xi(r)$  and $P(k)$  form a  Fourier  transform pair.
This is no longer true for  a non--Gaussian field for which $\xi$ must
be complemented  by other statistical descriptors  which are sensitive
to  the  structure  of  matter  on large  scales.   In  the  so-called
``standard  model''  of  structure  formation, an  initially  Gaussian
density  distribution becomes non--Gaussian  due to  mode-coupling and
the resultant  build up of  phase correlations during  the non--linear
regime.  These  phase correlations give rise to  the amazing diversity
of form, which  is characteristic of a highly  evolved distribution of
matter,  and is  often  referred to  as  being cellular,  filamentary,
sheet--like, network--like,  sponge--like, a cosmic web  etc.  Most of
these  descriptions  are  based  either  on  a  visual  appearance  of
large--scale structure or on the  presence of features that are absent
in  the  reference  Gaussian  distribution which,  by  definition,  is
assumed  to be featureless.   Gravitational instability,  for example,
may cause CDM--like Gaussian initial perturbations to evolve towards a
density field that percolates at  a higher density threshold, i.e.\ at
a lower filling factor, than a Gaussian field {\cite{melott:cluster}}.
Such  distributions  display greater  connectivity  and are  sometimes
referred to as being ``network--like'' {\cite{yess:universality}}.

In  order  to come  to  grips  with  the rich  textural  possibilities
inherent in large--scale structure, a number of geometrical indicators
of  clustering  have  been  proposed  in the  past  including  minimal
spanning   trees  {\cite{barrow:minimalspanning}},  the   genus  curve
{\cite{gott:sponge}},                percolation                theory
{\cite{zeldovich:maximum,shandarin:percolation}}  and  shape  analysis
(\pcite{sathyaprakash:morphology,sahni:approximation}  and  references
therein).    A  major   recent   advance  in   our  understanding   of
gravitational clustering  has been associated with  the application of
Minkowski functionals  (MFs) to cosmology  {\cite{mecke:robust}}.  The
four  MFs $V_0,\ldots,V_3$  provide  an excellent  description of  the
geometrical properties of a collection of point objects (galaxies) or,
alternatively, of  continuous distributions such as  density fields in
large--scale structure or brightness  contours in the cosmic microwave
background.  The scope and descriptive power of the MFs is enhanced by
the  fact that  both  percolation  analysis and  the  genus curve  are
members   of   the   family.    Additionally,   as   demonstrated   by
{\scite{sahni:shapefinders}},  ratios  of   MFs  provide  us  with  an
excellent ``shape  statistic'' with which one may  attempt to quantify
the  morphology of  large  scale structure,  including  the shapes  of
individual superclusters and voids.  Spurred  by the success of MFs in
quantifying  the geometrical properties  of large--scale  structure we
apply the MFs to scale--invariant N--body simulations of gravitational
clustering, with  an attempt to  probe both the global  properties and
the individual  ``bits and  pieces'' that might  make up  the ``cosmic
web'' {\cite{bond:filament}}.

\section{Method}

\subsection{Minkowski functionals}

Minkowski  functionals (named after  {\pcite{minkowski:volumen}}) were
introduced  into  cosmology  by {\scite{mecke:robust}}  employing  the
generalized Boolean grain model.  This  model associates a body with a
point  process (in  our  case given  by  the location  of galaxies  or
clusters) by  decorating each  point with a  ball of radius  $r$.  The
union  set of  the  covering balls  is  then studied  morphologically,
whereby  the radius  of  the  ball serves  as  a diagnostic  parameter
probing the spatial scale of the body.

In  this  paper we  shall  use the  excursion  set  approach which  is
applicable to  continuous fields (that  may be constructed  from point
processes).   {\scite{schmalzing:beyond}}  pointed  out how  to  apply
Minkowski functionals  to isocontours of continuous  fields, where the
contour  level (the  threshold) is  employed as  diagnostic parameter.
The excursion set approach inherits two diagnostic parameters, because
we  may  also  vary  the  smoothing scale  used  in  constructing  the
continuous field.

In three  dimensions there  exist four Minkowski  functionals $V_\mu$,
$\mu=0,1,2,3$.  They  provide a complete  and unique description  of a
pattern's  global  morphology  in  the  sense  of  Hadwiger's  theorem
{\cite{hadwiger:vorlesung}}.  While reducing the information contained
in  the full  hierarchy of  correlation functions,  this small  set of
numbers  incorporates correlations  of arbitrary  order  and therefore
provides  a   complementary  look  on   large--scale  structure.   The
geometric  interpretations  of  all  Minkowski  functionals  in  three
dimensions are summarized in Table~\ref{tab:minkowski}.

We calculate  global Minkowski functionals of  the isodensity contours
of  the density  field as  described  in Appendix~\ref{app:minkowski}.
Furthermore, we separately calculate the partial Minkowski functionals
of  each isolated  part of  the isodensity  contour.  Since  the total
isodensity  contour  is  the  union  of  all  its  parts,  the  global
functionals are given  as sums of the partial  functionals at the same
threshold.  This  follows the spirit  of {\scite{mecke:robust}}, where
partial  Minkowski   functionals  are  introduced   to  measure  local
morphology for the generalized Boolean grain model.

Partial Minkowski  functionals (PMF) offer the  possibility of probing
the morphology  of individual  objects, or the  object's environmental
morphological properties,  respectively.  We expect  that this concept
will  be more  powerful  when  applied to  continuous  fields at  high
spatial resolution so that the  details of structures are not smoothed
out.  Their  application to point processes also  delivers more direct
information.    PMF  provide   a  wide   range  of   possibilities  for
morphological studies which we shall explore in a forthcoming paper.

\subsection{Shapefinders}

One important task of morphological statistics consists in quantifying
strongly non--Gaussian features such as filaments and pancakes.  Given
the four Minkowski functionals, we aim at reducing their morphological
information content  to two  measures of planarity  and filamentarity,
respectively, as  has been done  for example with  various geometrical
quantities     {\cite{mo:statistical}},     moments     of     inertia
{\cite{babul:filament}},    and    cumulants   of    counts--in--cells
{\cite{luo:shape}}.

Recently,    {\scite{sahni:shapefinders}}    proposed    a   set    of
shapefinders  derived from Minkowski  functionals.  One  starts from
the  three  independent  ratios  of Minkowski  functionals  that  have
dimension  of length.   Requiring that  they yield  the radius  $R$ if
applied to a ball, we define
\begin{equation}
\text{Thickness }T:=\frac{V_0}{2V_1},\qquad
\text{Width }W:=\frac{2V_1}{{\pi}V_2},\qquad
\text{Length }L:=\frac{3V_2}{4V_3}.
\end{equation}
By  the isoperimetric  inequalities~(\ref{eq:isoperimetric}),  we have
$L{\ge}W{\ge}T$ for any convex body.

Going  one  step  further,  {\scite{sahni:shapefinders}}  also  define
dimensionless shapefinders by
\begin{equation}
\label{eq:shapefinders}
\text{Planarity }\cP:=\frac{W-T}{W+T},\qquad
\text{Filamentarity }\cF:=\frac{L-W}{L+W}.
\end{equation}

Some examples are in order.

\subsection{Simple examples}

Let   us  consider   some  simple   families  of   convex   bodies  in
three--dimensional  space that  can take  both filamentary  and planar
shape.

% Ellipse with radii $r$ and $\lambda{r}$ embedded in three--dimensional
% space\footnote{$E(k)=\int_0^{\pi/2}\rd\varphi\sqrt{1-k^2\sin^2\varphi}$
% is the complete elliptical integral of the second kind.}
% \begin{equation}
% V_0=0 ,\qquad
% V_1=\frac{\pi}{3}r^2\lambda ,\qquad
% V_2=\frac{2}{3}\rE(\sqrt{1-\lambda^2}) ,\qquad
% V_3=1
% \end{equation}

A  spheroid  with two  axes  of  length $r$  and  one  axis of  length
$\lambda{r}$ has Minkowski functionals
\begin{equation}
V_0= \frac{4\pi}{3}r^3\lambda ,\qquad
V_1= \frac{\pi}{3}r^2\left(1+f(1/\lambda)\right) ,\qquad
V_2= \frac{2}{3}r\left(\lambda+f(\lambda)\right) ,\qquad
V_3= 1,
\label{eq:spheroid}
\end{equation}
where\footnote{Note that for arguments $x>1$, one can use the relation
$i\arccos{x}=\ln\left(x+\sqrt{x^2-1}\right)$ to  recover an explicitly
real--valued expression.}  
\begin{equation}
f(x)=\frac{\arccos{x}}{\sqrt{1-x^2}}.
\end{equation}
By  varying the  parameter $\lambda$  from  zero to  infinity, we  can
change the morphology of the spheroid from a filament to a pancake via
a spherical cluster.

A  different way  of  deforming a  filament  into a  pancake goes  via
generic  triaxial ellipsoids;  thereby one  of the  smaller axes  of a
strongly prolate  spheroid is increased  in size until it  matches the
larger  axis and  an oblate  spheroid has  been reached.   An integral
expression can be found in {\cite{sahni:shapefinders}}.

Yet another  transition from  prolate to oblate  shape is  provided by
cylinders of radius $r$  and height $\lambda{r}$.  Here, the Minkowski
functionals are given by
\begin{equation}
V_0=\pi r^3\lambda ,\qquad
V_1=\frac{\pi}{3}r^2(1+\lambda) ,\qquad
V_2=\frac{1}{3}r(\pi+\lambda) ,\qquad
V_3=1.
\label{eq:cylinder}
\end{equation}

\begin{figure}[h]
\begin{center}
\epsfbox{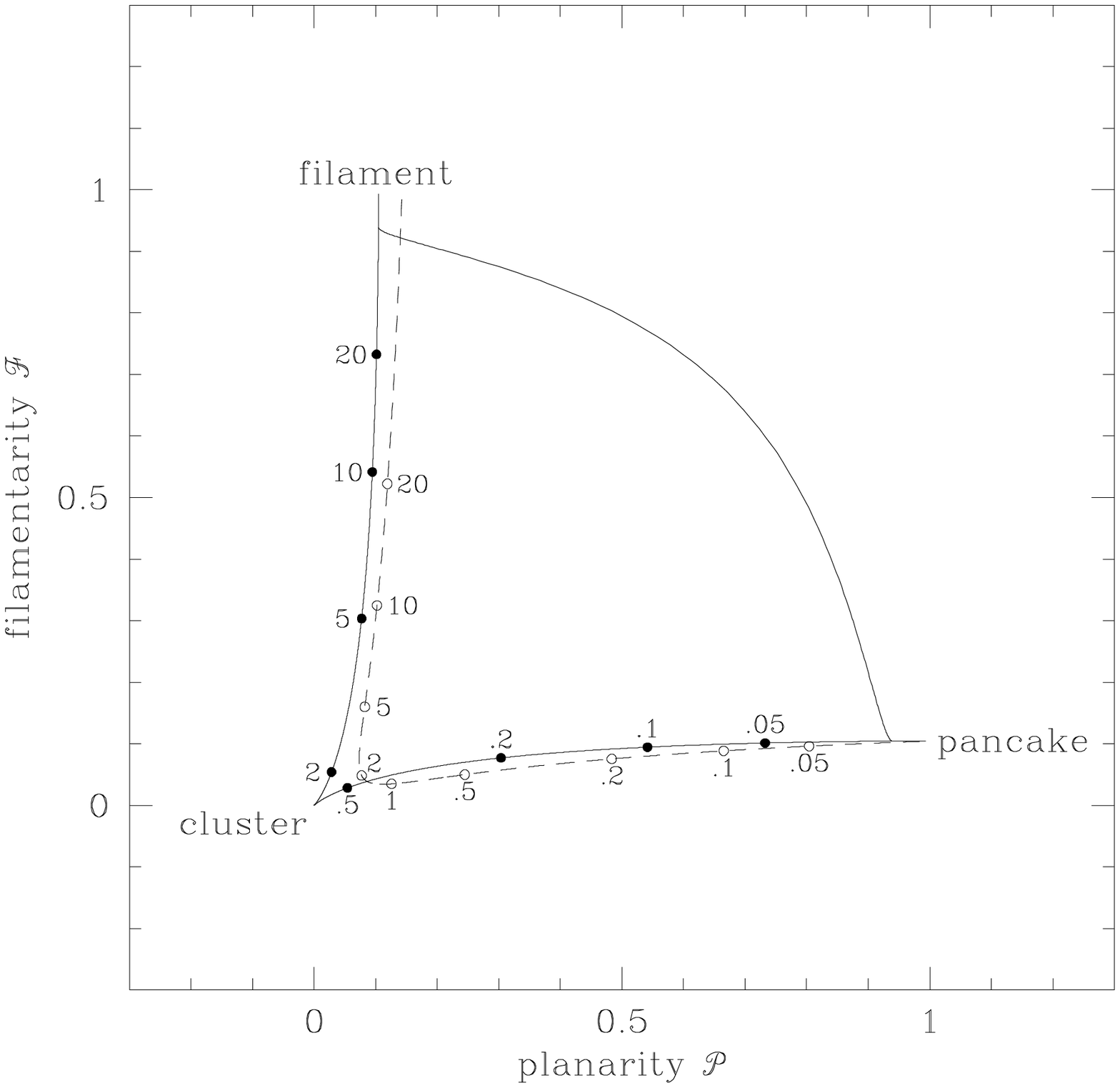}
\end{center}
\caption{
\label{fig:blaschke.simple}
The  shapefinders of  some convex  bodies.  The  solid line  shows the
transition from a ball  to a filament to a pancake and  back to a ball
via  triaxial   ellipsoids  of   various  shapes.   The   dashed  line
corresponds  to cylinders  that  also undergo  the  transition from  a
filament  to a  pancake  by varying  their  height.  The  dots on  the
spheroid and cylinder curves  indicate typical values of the $\lambda$
parameter      from      Equations~(\protect\ref{eq:spheroid})     and
(\protect\ref{eq:cylinder}), respectively. }
\end{figure}

A  Blaschke   diagram,  that  is   a  plot  of  planarity   $\cP$  and
filamentarity $\cF$, summarizing the morphological properties of these
simple convex bodies is shown in Figure~\ref{fig:blaschke.simple}.

\section{A set of $N$--body simulations}

\subsection{Description}

We start from a family of initial power law spectra $P(k)\propto{k}^n$
with $n\in\{-2,-1,0,+1\}$ set before an Einstein--de~Sitter background
($\Omega$=1, $\Lambda$=0).   We conduct numerical  experiments using a
PM code (consult  {\pcite{melott:controlled}} for details).  Four sets
of phases  were used for each  model, making a total  of 16 simulation
runs.  Each run consists of $128^3$ particles sampled at an epoch well
in the non--linear  regime.  This epoch is chosen  such that the scale
of  non--linearity $k_{\text{nl}}$,  defined in  terms of  the evolved
spectrum
\begin{equation}
\sigma^2_{k_{\text{nl}}} = \int_0^{k_{\text{nl}}}\rd^3k P(k) = 1,
\end{equation}
is equal to  eight in units of the fundamental  mode of the simulation
box.   By  using  the  stage  $k_{\text{nl}}=8$,  we  make  sure  that
structure is already sufficiently developed on scales much larger than
the  simulation's  resolution,  while  it  is not  yet  influenced  by
boundary effects.

Using  a  cloud--in--cell kernel,  these  particles  were  put onto  a
$256^3$ grid, which  is the maximum value an  ordinary workstation can
tackle with acceptable time and memory consumption.  Subsequently, the
density    field    was    smoothed    with    a    Gaussian    kernel
$\propto\exp\left(-x^2/2\lambda^2\right)$,  where $x$ is  the distance
in mesh units  and the width $\lambda$ is set to  3.  Tests have shown
that this value both leads to a reasonably smooth field, and preserves
at least  some detail on  smaller scales.  Throughout the  article, we
re--scale the  density to the density contrast  $\delta$, ranging from
$-1$ to infinity with zero mean.

\subsection{The global field}

The global  Minkowski functionals  calculated from the  density fields
described     in    the    previous     section    are     shown    in
Figure~\ref{fig:epoch.f}.  The  four different line  styles correspond
to the  different spectral indices.  Figure~\ref{fig:epoch.f.rescaled}
shows  the Minkowski  functionals  for  the same  set  of models,  but
instead  of the  density  threshold $\delta$,  the rescaled  threshold
$\nu$ is used  as the $x$--axis.  $\nu$ is  calculated from the volume
Minkowski  functional  $v_0$,  that  is  the filling  factor  $f$,  as
described by {\scite{gott:quantitative}}.  Essentially, its use forces
exact  Gaussian  behavior  of   the  volume  $v_0$,  by  the  implicit
connection
\begin{equation}
v_0(\delta)=\frac{1}{2}-\frac{1}{2}\Phi\left(\frac{\nu}{\sqrt{2}}\right).
\end{equation}
Thus the  deviations from Gaussianity that  are due to  changes in the
one--point   probability  distribution   function  are   removed,  and
deviations due to higher--order correlations are emphasized.

Obviously,  the global  functionals clearly  discriminate  between the
various  models.   However,  in  order  to make  this  statement  more
quantitative, let  us take  a closer look  at the  individual coherent
objects composing the isodensity contours.

\subsection{The largest objects}

At  intermediate thresholds,  the excursion  sets consist  of numerous
isolated objects.  We identify them by grouping adjacent occupied grid
cells  into one object,  where adjacent  means that  the cells  have a
common face.  Since the Minkowski  functionals of the global field are
calculated  by integrating  over quantities  that can  be  assigned to
individual  grid  cells, the  partial  Minkowski  functionals of  each
object can  be obtained at no  extra cost once the  cells belonging to
each object have been identified.

\begin{figure}
\epsfbox{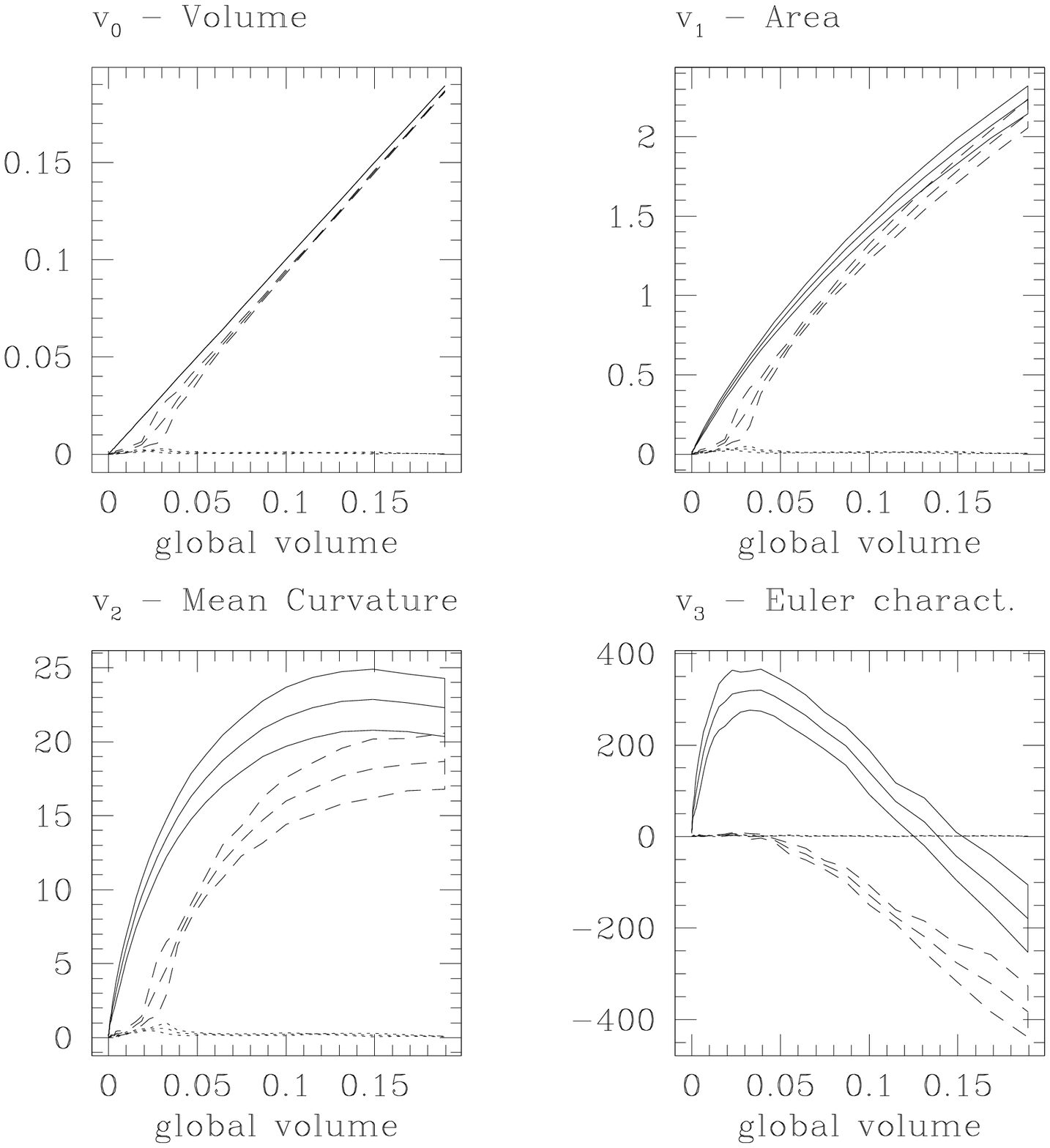}
\caption{
\label{fig:clusters.abcd}
The following set of figures compares the global Minkowski functionals
of the whole  density field (solid line) to  the Minkowski functionals
of  the largest  (dashed) and  second largest  (dotted)  objects.  The
areas  indicate  the  mean   and  standard  deviation  over  the  four
realizations of each  model.  In order to emphasize  the region around
the  percolation  transition, the  Minkowski  functionals are  plotted
against the filling  factor, which is equal to  the volume enclosed in
all isocontours  of the  global field.  In  this particular  plot, the
values for  an initially power--law  spectrum with an index  of $n=-2$
are shown.  }
\end{figure}

\begin{figure}				
\epsfbox{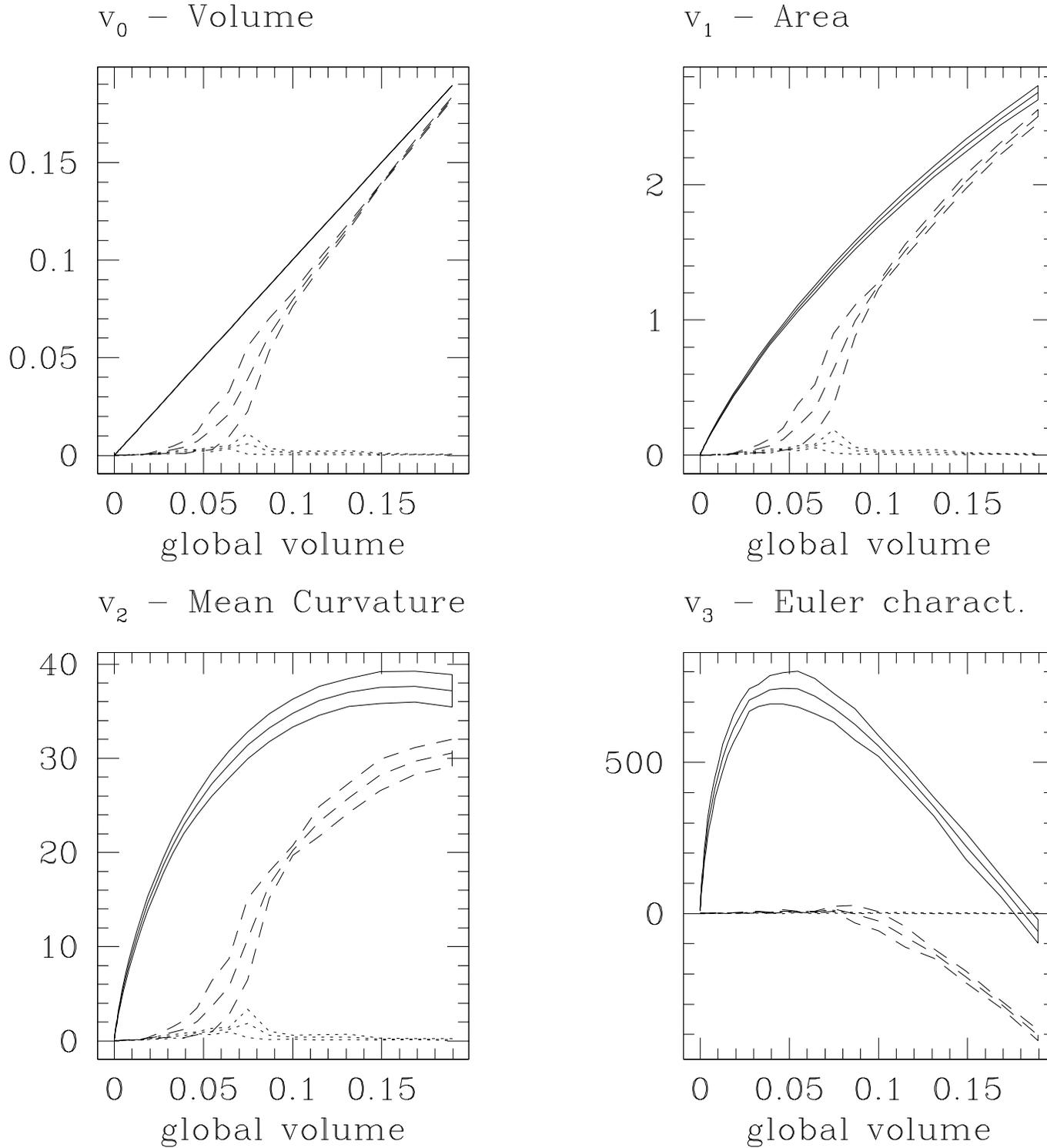}
\caption{
\label{fig:clusters.ijkl}
The same tendency as before is  visible in this case where $n=-1$.  As
before, the dominance of large networked structures is clearly visible
-- for example,  the Euler characteristic  stays well below  zero even
for  high thresholds,  which indicates  a spongy  structure  with many
tunnels in  it.  However, the  networked structures have  fewer loops,
since they tend to split up at lower thresholds.  }
\end{figure}

\begin{figure}				
\epsfbox{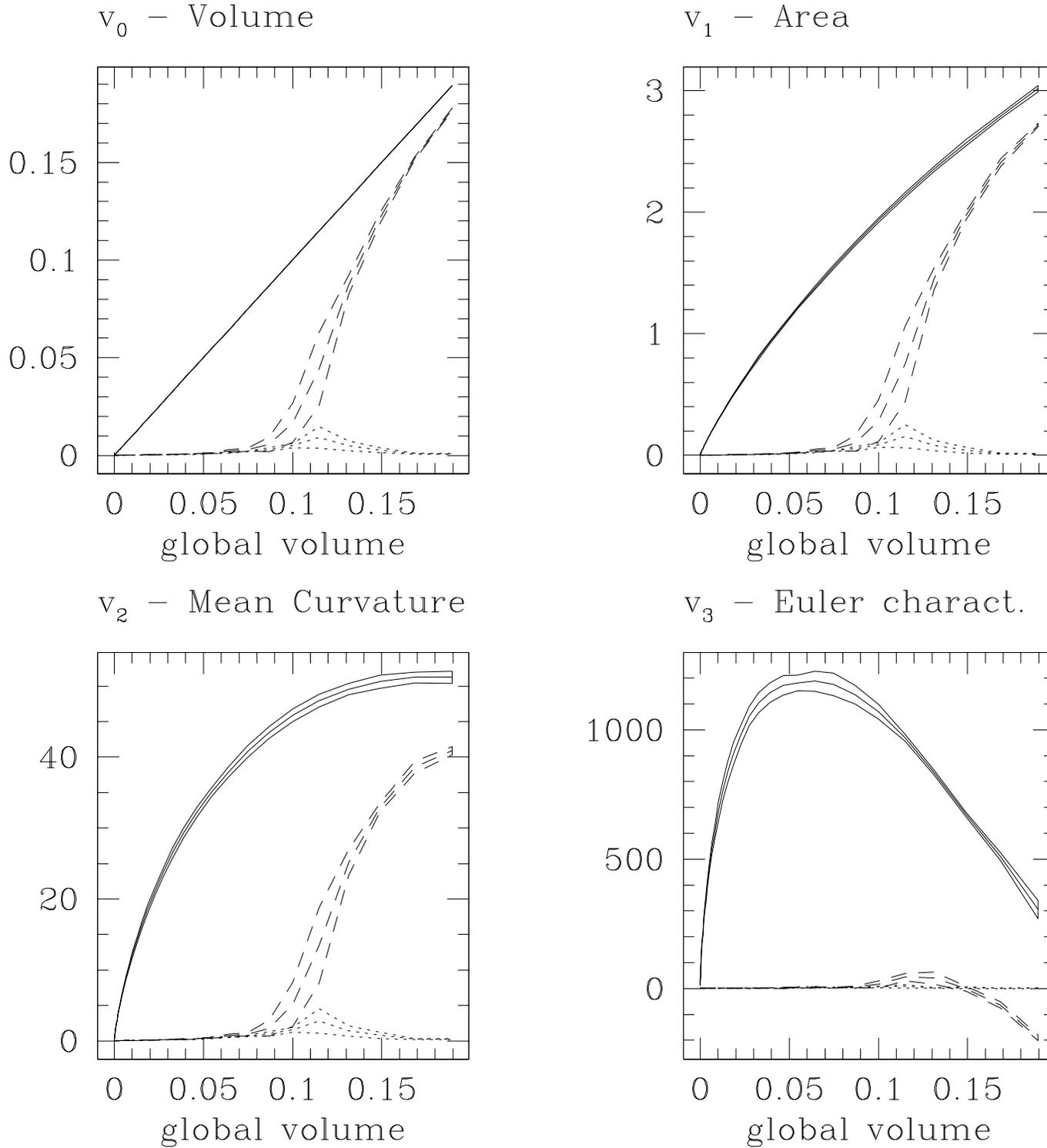}
\caption{
\label{fig:clusters.uvwx}
The  situation is  different in  the case  $n=0$.  The  largest object
splits  up rapidly,  and is  already  simply connected  at fairly  low
density  contrast ($\delta\approx0.5$).   Also,  at thresholds  around
percolation,  the  second largest  object  becomes  comparable to  the
largest  one, so  the field  is not  dominated by  a  single networked
structure, but by several chunks.  }
\end{figure}

\begin{figure}				
\epsfbox{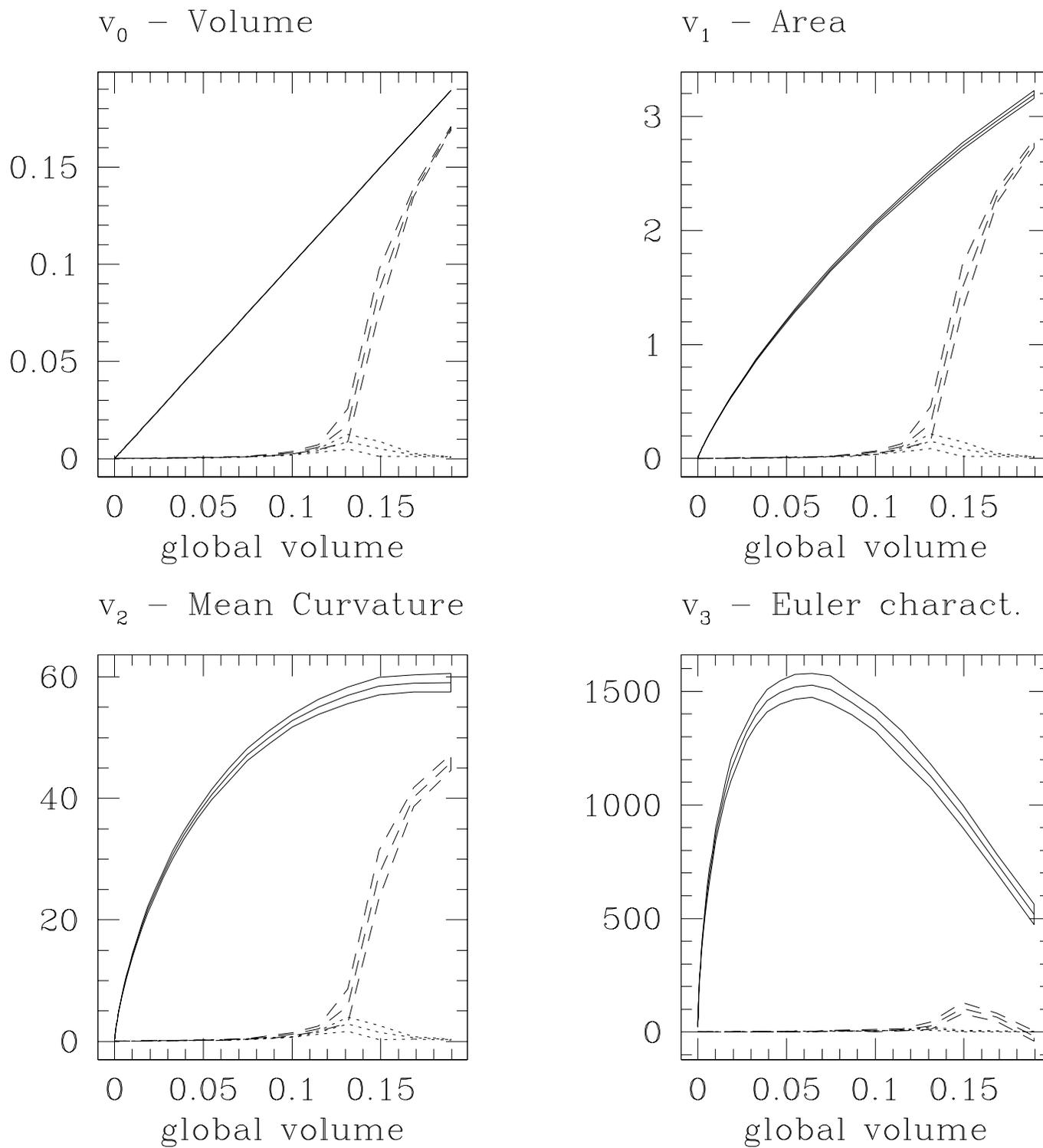}
\caption{
\label{fig:clusters.qrst}
The effect  described in the  previous figure is even  more pronounced
for $n=+1$, since this  model is dominated by small--scale structures.
Contrary  to the  models  with  negative index  $n$,  where the  Euler
characteristic of  the largest object  approaches the value one  for a
simply  connected structure  from  below,  one can  even  see a  small
positive peak around the percolation threshold.  }
\end{figure}

Several        plots        in        Figures~\ref{fig:clusters.abcd},
{\ref{fig:clusters.ijkl}},        {\ref{fig:clusters.uvwx}},       and
{\ref{fig:clusters.qrst}}  illustrate the  behavior  of the  Minkowski
functionals of these objects.  

Obviously, the contribution of smaller objects to the volume is almost
negligible compared to the largest one.  Note that in all figures, the
mean  and standard  deviation  over all  four  realizations are  shown
instead  of  the individual  curves.   It  is  worth noting  that  the
variance is largest in the $n=-2$ and $-1$ models, which are dominated
by  structures on  large scales  and hence  show the  strongest sample
variance.

The   models  with   various   initial  spectral   indices  $n$   show
qualitatively  a similar  behavior.   At small  filling factors  (high
density   thresholds)  the  two   largest  clusters   give  negligible
contribution  to  each  of   the  global  characteristics.   Then,  at
percolation  transition  the   largest  cluster  quickly  becomes  the
dominant  structure in  terms  of volume,  area,  and integrated  mean
curvature.  The  second largest cluster also grows  at the percolation
threshold  but  just  a  little  and  then  quickly  diminishes.   The
percolation transition is clearly  marked in all three characteristics
of  the  largest  cluster  by  their  sudden  growth.   However,  this
transition  does not  happen at  a well--defined  threshold.  Instead,
clusters gradually merge into the  largest objects as the threshold is
decreased (the filling factor  grows).  This continuous transition has
also  been  observed  using  percolation analysis,  i.e.\  the  zeroth
Minkowski                       functional                       alone
{\cite{shandarin:topology,shandarin:detection,klypin:percolation,sahni:probing}}
and is explained by the finite size of the sample.

Nevertheless, the percolation  transitions happen within fairly narrow
ranges of the  filling factor that are clearly  distinct for different
models   in   question:   The   filling  factors   are   approximately
$0.03\pm0.01,0.07\pm0.015,0.11\pm     0.015,0.14\pm0.015$    in    the
$n=-2,-1,0,+1$  models,  respectively.   It  is  remarkable  that  all
percolation  transitions  occur at  smaller  filling  factors than  in
Gaussian  fields  (about  0.16)  indicating  that  even  in  the  most
hierarchical model  ($n=+1$) the structures tend to  be more connected
than      in      the      ``structureless''      Gaussian      field.
{\scite{pauls:hierarchical}} showed positive correlation with networks
based on  the same phases  all the way  to $n=+3$.  This  confirms the
conclusion  of {\scite{yess:universality}}:  The  universality of  the
network  structures results  from  the evolution  of Gaussian  initial
conditions    through    gravitational    instability.    The    Euler
characteristic  of  the largest  cluster  also  marks the  percolation
threshold but in a different manner: before percolation it is zero and
after percolation it becomes negative  in every model, however, in the
$n=0$  and $n=+1$  models it  grows to  a small  positive  peak before
becoming negative.  All global functionals have no particular features
at the percolation threshold.

\subsection{Small objects}

\begin{figure}
\epsfbox{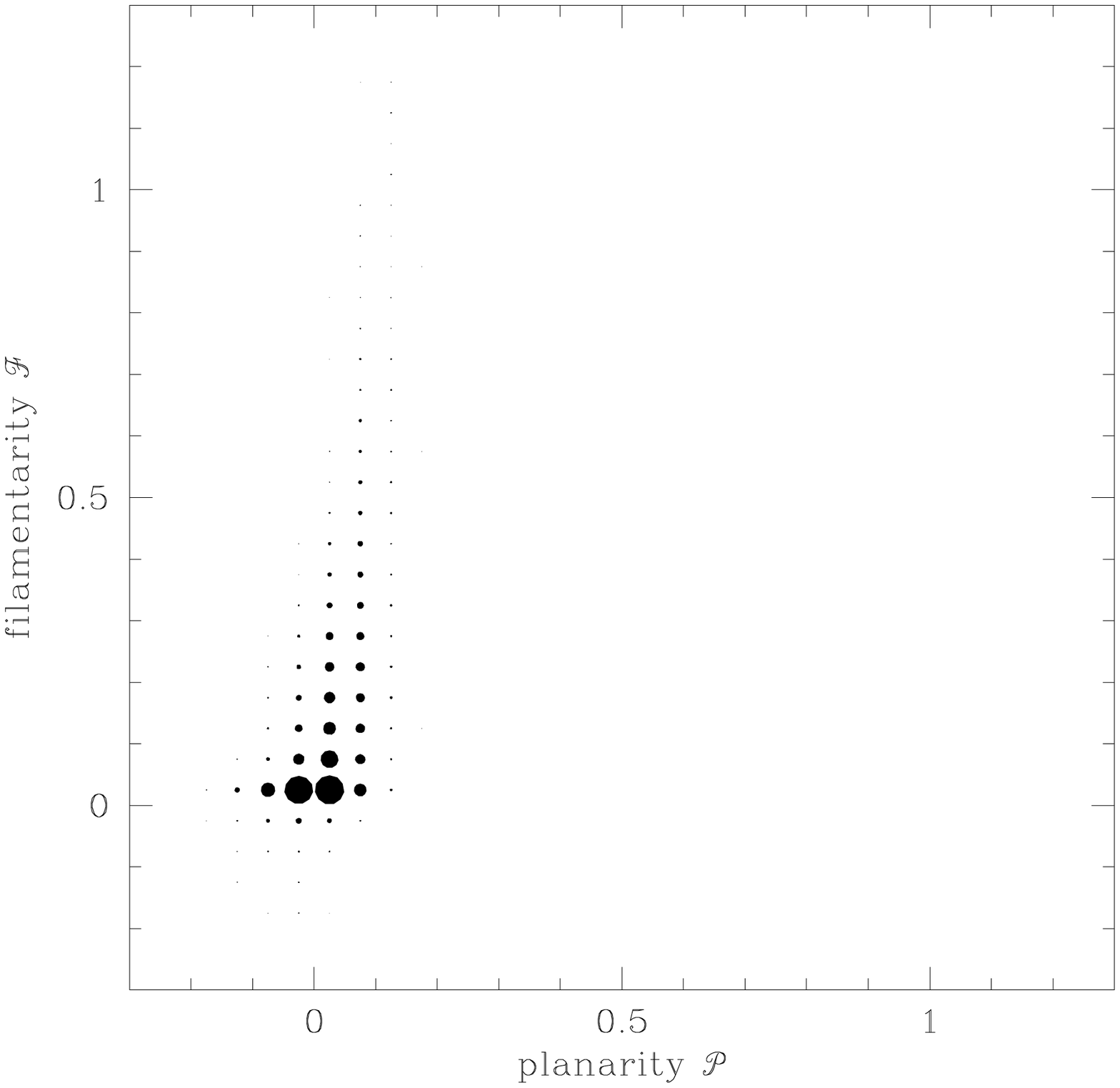}
\caption{
\label{fig:blaschke.abcd}
The shapefinder  or Blaschke  diagram of the  model with  $n=-2$.  The
area of each  dot corresponds to the number  of coherent objects whose
shapefinders $(\cP,\cF)$ lie around the center of this dot.  Note that
the  shapefinders of all  objects at  all thresholds  are used  in the
construction  of the  diagram.  Nevertheless,  most  information comes
from  thresholds  close  to   percolation,  where  small  objects  are
abundant. }
\end{figure}

As an example,  the Blaschke diagram for the model  $n=-2$ is shown in
Figure~\ref{fig:blaschke.abcd}. The distributions for the other models
look qualitatively  very similar and the average  quantities for other
models     are    presented     in     Figures~\ref{fig:mass.x}    and
{\ref{fig:mass.y}}.   Figure~\ref{fig:mass.x} shows  that most  of the
small  objects are either  spherical or  slightly planar  (two largest
dots in Figure~\ref{fig:blaschke.abcd}).  There is also a considerable
number of elongated clusters with  filamentarities from 0.1 to 0.5. In
some  cases  filamentarity  reaches  large  values  $\sim1$.   On  the
contrary planarity is much weaker,  it hardly reaches the value of 0.2
(which is partly a consequence of  the smoothing).  There is a hint of
a small  correlation between filamentarity and  planarity: the objects
with the largest filamentarity also tend to have larger planarity.

Figures~\ref{fig:mass.x}    and    {\ref{fig:mass.y}}   display    the
shapefinders  as  functions of  the  cluster  mass.   The curves  give
averages over the realizations  of each model.  Apparently, the signal
for filamentarity  is much stronger than for  planarity, regardless of
the     model     which      is     in     full     agreement     with
Figure~\ref{fig:blaschke.abcd}.    The  planarity   and  filamentarity
distributions  qualitatively look very  similar except  the amplitude.
Small    objects   ($5\times10^{-6}\lesssim{m}\lesssim5\times10^{-4}$)
display  stronger planarity  and  filamentarity for  models with  more
power    on    large   scales.     However,    for   greater    masses
($m\gtrsim5\times10^{-4}$)  the   situation  is  reversed:   the  less
large-scale power the greater  the filamentarity and planarity. If the
former  seems to  be natural  and was  expected, the  latter  has been
unexpected.  Both  the planarity and  filamentarity monotonically grow
and  reach their  maxima  at largest  clusters: $\cP_m\approx0.1$  and
$\cP_m\approx0.5$  in all  models.   As expected  the largest  objects
possess   the  largest  planarities   and  filamentarities,   but  the
independence  of  the maxima  from  the  model  again was  unexpected.
Figures~\ref{fig:blaschke.x}   and  {\ref{fig:blaschke.y}}   show  the
histograms for  the shapefinders $\cP$ and  $\cF$, respectively.  They
clearly show the  large difference in the total  number of structures:
the more power on small  scales the greater the abundance of clusters.
These    figures     are    in    general     agreement    with    the
Figures~\ref{fig:mass.x} and {\ref{fig:mass.y}}.

\section{Conclusion and Outlook}

\begin{figure}[h]
\epsfbox{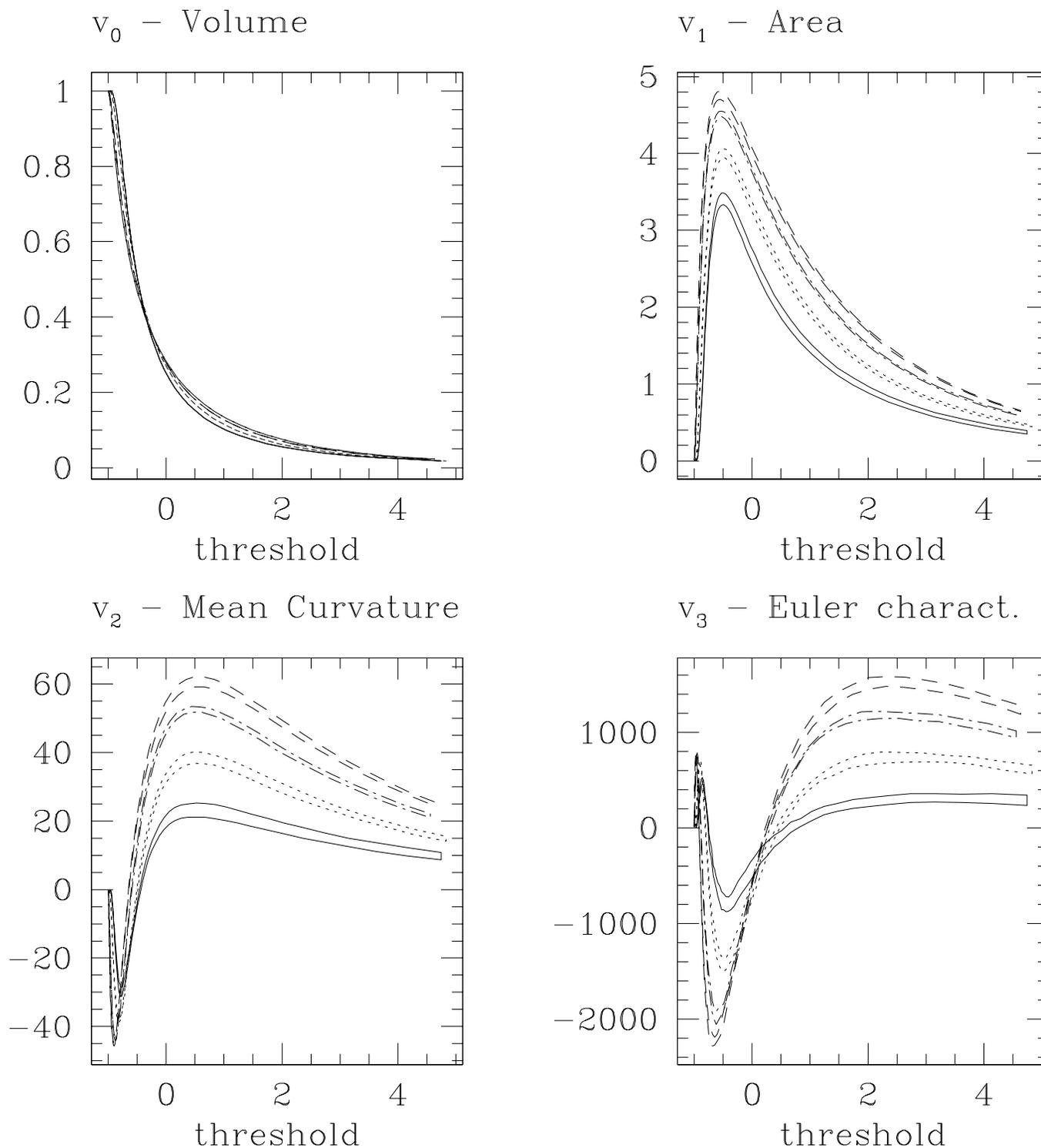}
\caption{
\label{fig:epoch.f}
Minkowski functionals for  the evolutionary stage $k_{\text{nl}}=8$ of
four    different   models    with   initial    power    law   spectra
$P(k)\propto{k}^n$.  The  area between two lines  gives the one--sigma
deviations  of four  realizations from  their mean.   The  solid lines
correspond to  $n=-2$, dotted is  $n=-1$, dash--dotted lines  give the
results for $n=0$, and $n=+1$ is shown with dashed lines.  }
\end{figure}

\begin{figure}[h]
\epsfbox{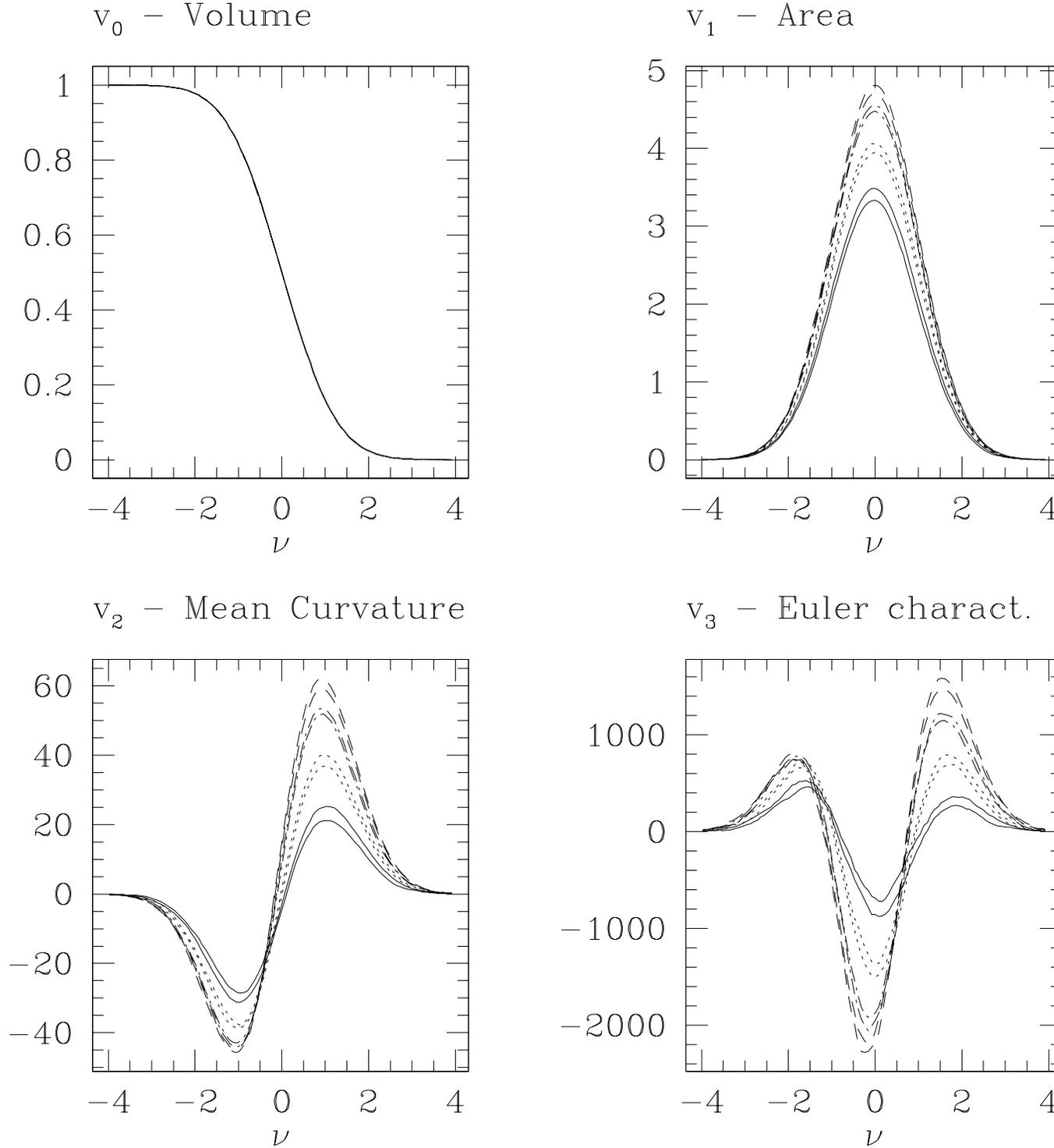}
\caption{
\label{fig:epoch.f.rescaled}
Minkowski functionals for  the evolutionary stage $k_{\text{nl}}=8$ of
four different models with initial power law spectra.  Line styles are
the  same  as in  the  previous figure,  but  instead  of the  density
threshold  $\delta$,  the   rescaled  threshold  $\nu$  introduced  by
{\protect\scite{gott:quantitative}} is used.  }
\end{figure}

\begin{figure}
\epsfbox{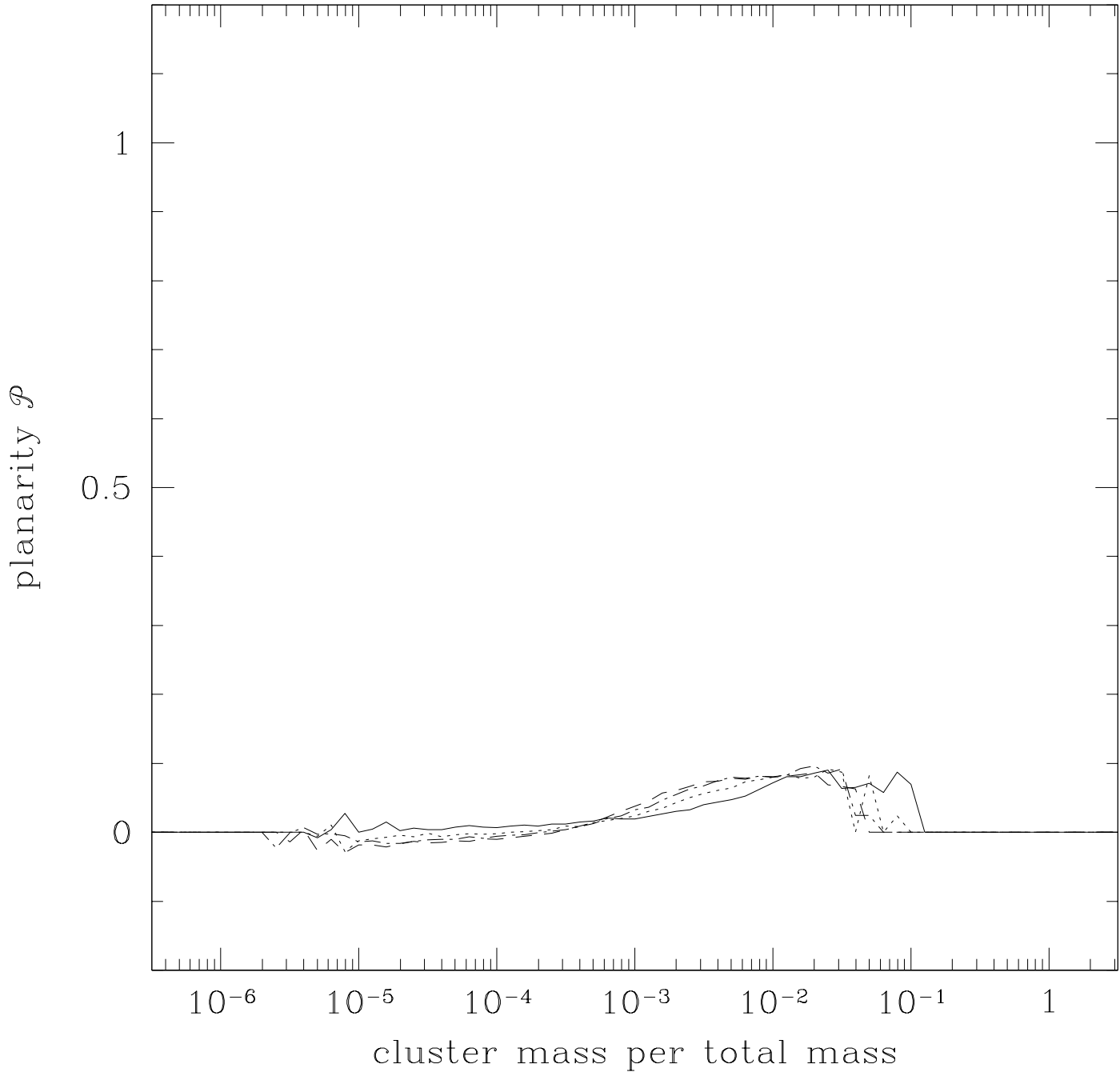}
\caption{
\label{fig:mass.x}
The average planarity  shapefinder $\cP$ as a function  of the cluster
mass.  Line styles  are explained in Figure~\protect\ref{fig:epoch.f}.
As  in  Figure~\protect\ref{fig:blaschke.abcd},  all  objects  at  all
thresholds are used to improve statistics.  }
\end{figure}

\begin{figure}
\epsfbox{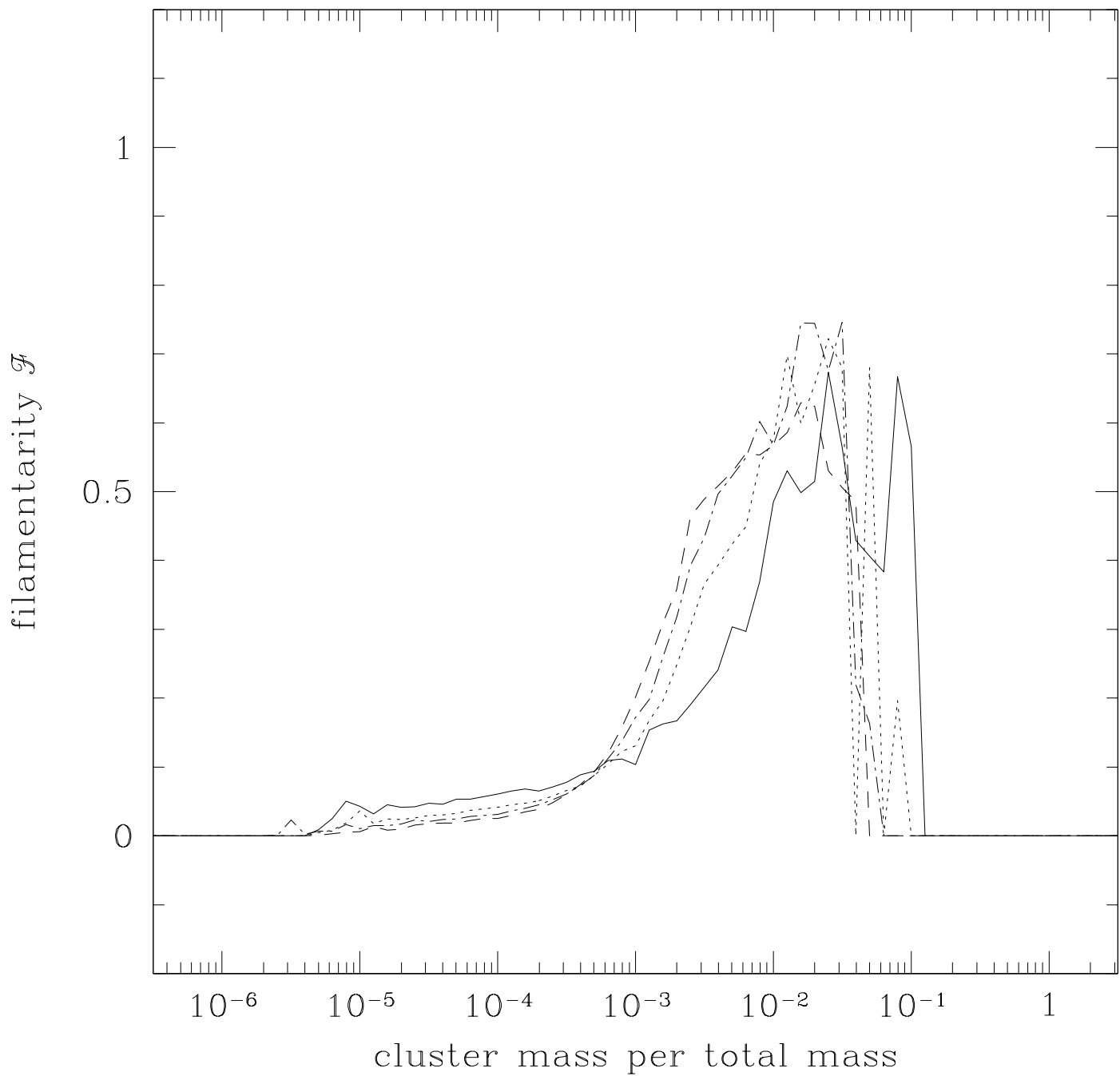}
\caption{
\label{fig:mass.y}
This  plot  shows  the  same  type of  distribution  as  the  previous
Figure~\protect\ref{fig:mass.x},    but   this   time    the   average
filamentarity  shapefinder  $\cF$ is  plotted  as  a  function of  the
cluster mass.  Again, line styles correspond to models as explained in
the caption of Figure~\protect\ref{fig:epoch.f}.  }
\end{figure}

\begin{figure}				
\epsfbox{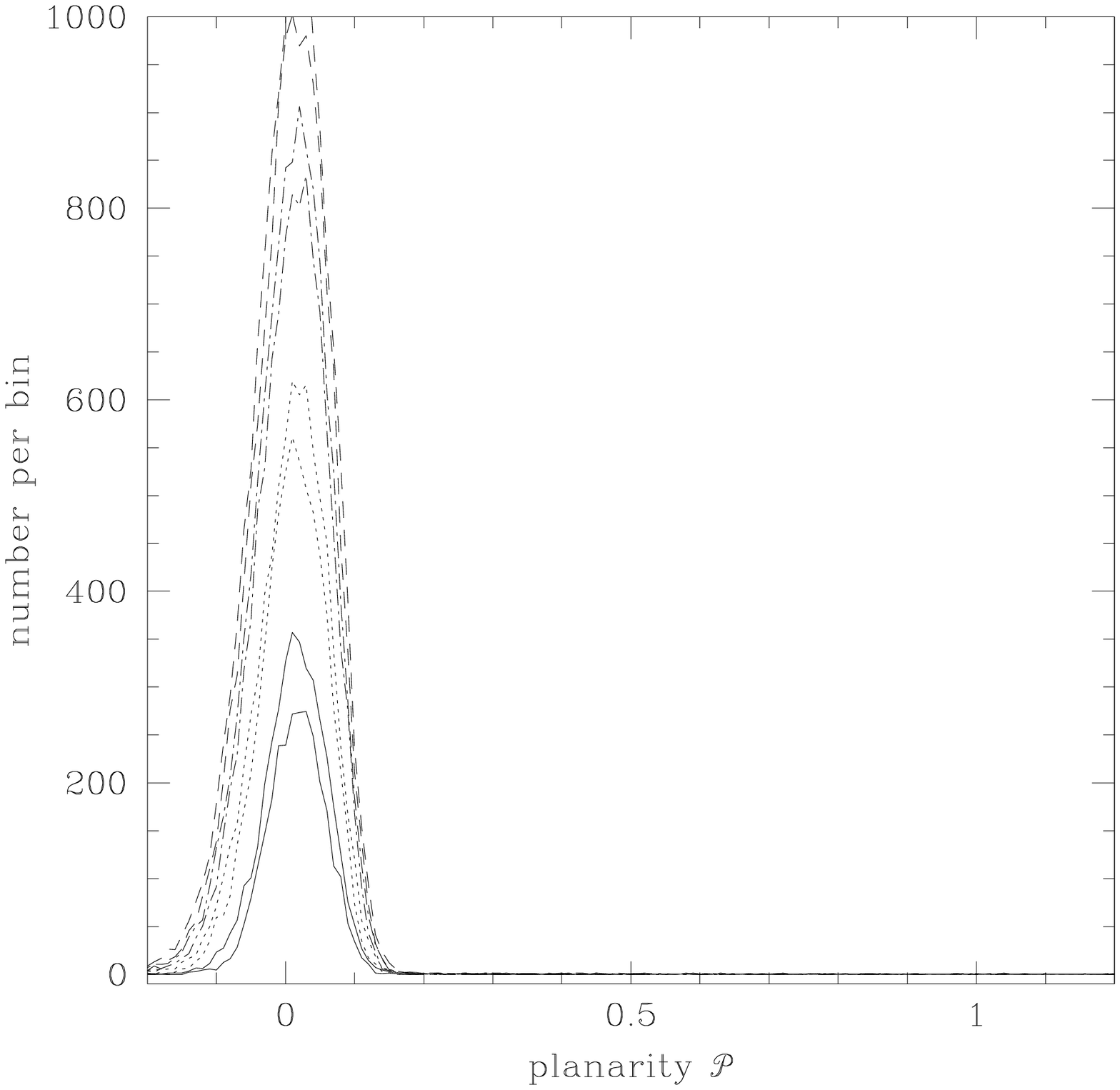}
\caption{
\label{fig:blaschke.x}
The distribution  of the  planarity shapefinder $\cP$.   The histogram
only includes objects larger  than $10^{-5}$ times the simulation box,
a  value of the  order of  the smoothing  volume.  Note  that although
theoretically,  the value  should  lie between  0  and 1,  measurement
errors  lead to  negative values.   The curves  give total  numbers of
objects  per   histogram  interval,   hence  they  differ   widely  in
normalization, but still agree  reasonably well in shape.  Line styles
are explained in Figure~\protect\ref{fig:epoch.f}.  }
\end{figure}

\begin{figure}				
\epsfbox{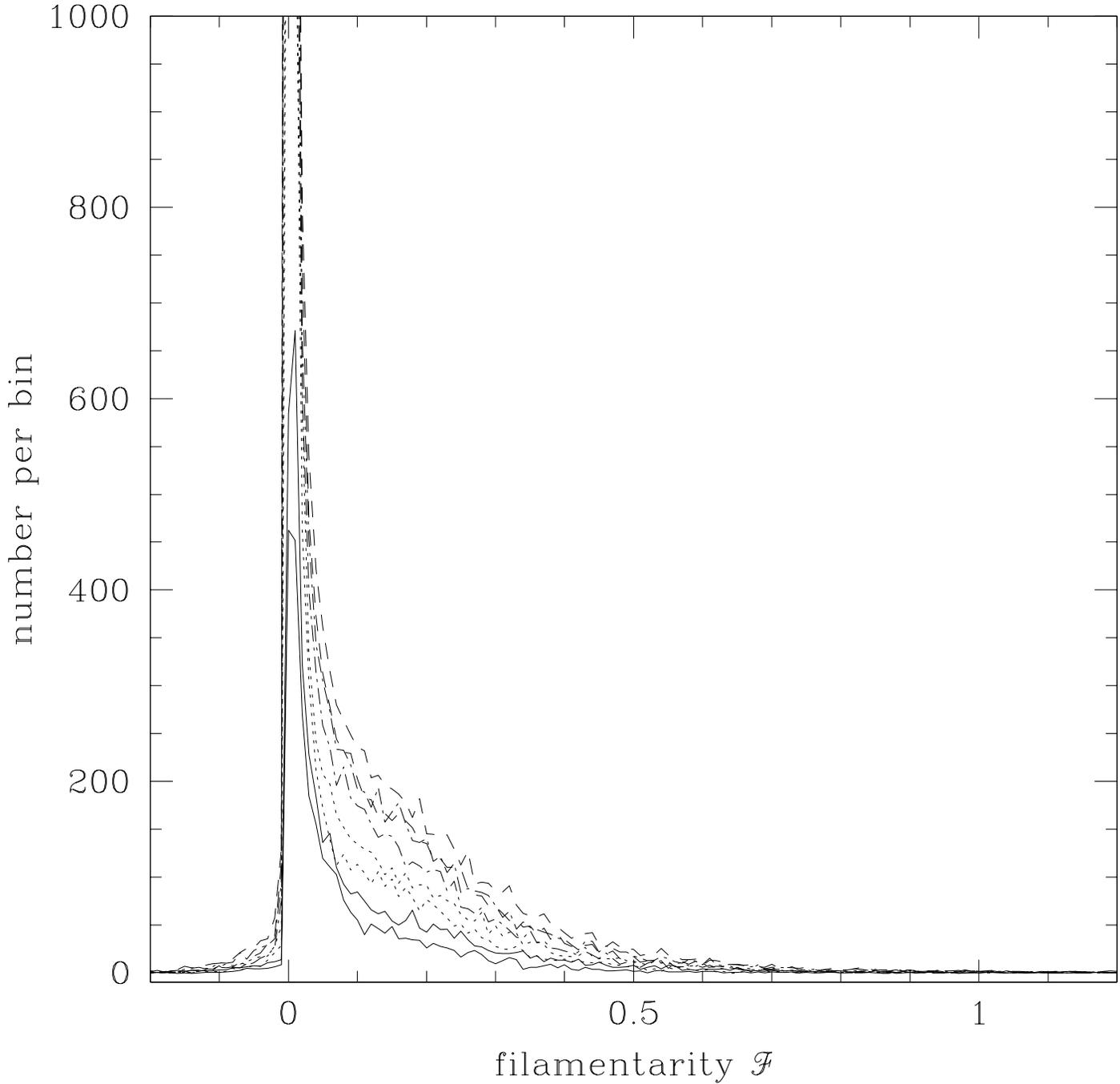}
\caption{
\label{fig:blaschke.y}
This figure displays the same type of distribution as above, but shows
the filamentarity  shapefinder $\cF$.  Line styles  are explained in
Figure~\protect\ref{fig:epoch.f}.  }
\end{figure}

Global        Minkowski       functionals        do       discriminate
(Figures~\ref{fig:epoch.f}  and  {\ref{fig:epoch.f.rescaled}}).   Only
the  $n=-2$ model  shows  slight  drawbacks as  far  as robustness  is
concerned, but that is due to the method's sensitivity to large--scale
features  of  the  smoothed  density  field.   The  total  area,  mean
curvature, and Euler characteristic are sensitive to abundances of the
structures and are  easy to interpret: the more  power on small scales
(greater $n$)  the more abundant structures and  therefore the greater
the amplitude of the curve.

Even  more  valuable  information  is  obtained from  looking  at  the
Minkowski functionals of the largest coherent object at each threshold
(Figures~\ref{fig:clusters.abcd},            {\ref{fig:clusters.ijkl}},
{\ref{fig:clusters.uvwx}},  and  {\ref{fig:clusters.qrst}}). All  four
Minkowski  functionals  of the  largest  cluster clearly  consistently
detect the percolation transition.  Two points are worth stressing: 1)
in all  models the percolation  transition happens at  smaller filling
factors than in the ``structureless''  Gaussian fields and 2) the more
power  on the  large scales  (i.e. the  smaller $n$)  the  smaller the
filling factor at percolation. Both conclusions confirm the results of
{\scite{yess:universality}}  about  the  universality of  the  network
structures  in the  power law  models  with $n\le1$.   The results  of
{\scite{pauls:hierarchical}}  present  evidence  that this  should  be
expected all  the way  up to $n=+3$;  at $n=+4$ mode  coupling effects
from  smaller  scales  should  begin  to  fully  disrupt  the  network
structure.

Small  objects, on  the  other hand,  give  different results.   Their
abundance  discriminates  well,  but  is  already  determined  by  the
difference in  the Euler characteristic as  well as by  the total area
and mean curvature of the whole contour (Figures~\ref{fig:epoch.f} and
{\ref{fig:epoch.f.rescaled}}) that are also sensitive to the abundance
of  structures.   The  morphology  of  small objects  as  measured  by
shapefinders   shows  little   differences  between   models   so  far
(Figures~\ref{fig:mass.x},  {\ref{fig:mass.y}}, {\ref{fig:blaschke.x}}
and  {\ref{fig:blaschke.y}}).   Both  the  maximum  average  planarity
($\cP\approx0.1$)    and    the    maximum    average    filamentarity
($\cF\approx0.5$)  are  reached in  the  most massive  non-percolating
objects.  None of the  model showed ribbon--like objects characterized
by both large planarity and large filamentarity.  We may speculate that
the smaller objects are ones which formed earlier, are more nonlinear,
and therefore more decoupled from initial conditions.

However, all models used Gaussian initial conditions, and evolve under
the influence of gravity.  Hence  the similar morphology of the clumps
may point towards universal behavior.  Note that things such as string
wakes might produce totally different results.

The grouping and measurement techniques used in this study may be less
accurate  for small  objects than  for  large clusters.   It is  worth
trying  to study  the morphology  of  small objects  by applying  more
accurate methods of measuring  partial Minkowski functionals such as a
Boolean  grain model (\pcite{schmalzing:diplom},  and a  follow--up to
this     article)      or     the     interpolation      method     of
{\scite{novikov:minkowski}} generalized to three dimensions.

\section*{Acknowledgments}

JS  wishes to thank  Martin Kerscher  for interesting  discussions and
valuable  comments.    This  work  is   part  of  a  project   in  the
``Sonderforschungsbereich  375--95 f\"ur Astroteilchenphysik''  of the
Deutsche Forschungsgemeinschaft.   ALM and SFS  acknowledge support of
the  NSF--EPSCoR program  and the  GRF  program at  the University  of
Kansas.  SFS acknowledges support from TAC in Copenhagen.

\appendix

\section{Calculating Minkowski functionals of isodensity contours }
\label{app:minkowski}

Using  methods   from  integral  geometry   and  scale--space  theory,
{\scite{schmalzing:beyond}}  developed numerical  methods  to quantify
the morphology of isodensity contours of a random field.  They present
two complementary possibilities for estimating Minkowski functionals.

Given a  density field sampled at  the grid points of  a cubic lattice
with lattice  constant $a$, Crofton's  formula {\cite{crofton:theory}}
requires  counting  of  the  number  of grid  cells  per  unit  volume
contained in the excursion set.  We obtain
\begin{eqnarray}
\widehat{V}_0(\nu) &= n_3, \\
\widehat{V}_1(\nu) &= \frac{2}{9a} (n_2-3n_3), \\
\widehat{V}_2(\nu) &= \frac{2}{9a^2}(n_1-2n_2+3n_3), \\
\widehat{V}_3(\nu) &= \frac{1}{a^3} (n_0-n_1+n_2-n_3),
\end{eqnarray}
where  the quantity $n_j$  is the  number density  of $j$--dimensional
elementary cells; to be specific, $n_3$ is the number of grid volumes,
$n_2$ counts the grid faces, and $n_1$ and $n_0$ denote the numbers of
grid edges and lattice points, respectively.

Alternatively,  it is sufficient  to estimate  the derivatives  of the
random  field  at  the  grid  points and  perform  a  spatial  average
$\langle\cdot\rangle_\cD   =  \frac{1}{\cD}\int\limits_\cD\rd^3x\cdot$
over   Koenderink  invariants\footnote{The   notation  is   chosen  to
emphasize the coordinate invariance  of the integrand.  We use indices
following  a comma  to indicate  differentiation with  respect  to the
corresponding coordinate.  $\epsilon_{ijk}$  denotes the components of
the   totally   antisymmetric   third--rank   tensor   normalized   to
$\epsilon_{123}=1$.  Summation over pairwise indices is understood.}
\begin{eqnarray}
\widehat{V}_0(\nu) &= \left\langle\Theta(u-\nu)\right\rangle_\cD ,\\
\widehat{V}_1(\nu) &= \left\langle\frac{1}{6}\delta(u-\nu)(u_{,i}u_{,i})^{1/2}\right\rangle_\cD, \\
\widehat{V}_2(\nu) &= \left\langle\frac{1}{3\pi}\delta(u-\nu)\frac{\epsilon_{ijm}\epsilon_{klm}u_{,i}u_{,jk}u_{,l}}{2u_{,n}u_{,n}}\right\rangle_\cD, \\
\widehat{V}_3(\nu) &= \left\langle\frac{1}{4\pi}\delta(u-\nu)\frac{\epsilon_{ijk}\epsilon_{lmn}u_{,i}u_{,l}u_{,jm}u_{,kn}}{2(u_{,p}u_{,p})^{3/2}}\right\rangle_\cD.
\end{eqnarray}

Careful analysis  reveals that both families of  estimators are biased
by  the finite  lattice constant  $a$ encountered  with  any practical
realization of a random field.   The deviations are of order $a^2$ and
can  be  evaluated  analytically  for  a Gaussian  random  field  (see
{\pcite{winitzki:minkowski}} for some results).  It turns out that the
two  estimates  using  Crofton's  formula and  Koenderink  invariants,
respectively,  deviate from  the  true value  in opposite  directions.
Hence  their difference  gives at  least a  rough idea  of  the errors
associated with binning the continuous field onto a finite lattice.

Yet another method of calculating Minkowski functionals of isocontours
was   suggested   by   {\scite{novikov:minkowski}}.   However,   their
interpolation  method  has  so   far  only  been  implemented  in  two
dimensions.

\section{Isoperimetric inequalities and shapefinders }
\label{app:isoperimetric}

In  his original  article, {\scite{minkowski:volumen}}  introduced the
mixed volumes $V(K_1,\ldots,K_n)$ of $n$ convex bodies $K_1$ to $K_n$.
In terms of ordinary volumes of Minkowski sums of the bodies, we have
\begin{equation}
V(K_1,\ldots,K_n) := \frac{1}{n!} \sum_{k=1}^n (-1)^{n+k}
\sum_{i_1<\ldots<i_k} V(K_{i_1}\oplus\ldots\oplus{K}_{i_k}),
\end{equation}
where the Minkowski sum is defined as
\begin{equation}
K\oplus{L} := \left\{ x+y | x\in{K}, y\in{L} \right\}.
\end{equation}
In the case of a single body, the mixed volume reduces to its ordinary
volume.   Furthermore,  these   quantities  inherit  many  interesting
properties  from the  volume  and hence  play  a central  role in  the
Brunn--Minkowski   theory   (see   {\pcite{schneider:brunn}}  for   an
introduction).

A  most  useful  inequality  proved by  {\scite{alexandrov:neue}}  and
{\scite{fenchel:inegalites}} states that
\begin{equation}
V^2(K_1,K_2,\ldots,K_n) \geq
V(K_1,K_1,K_3\ldots,K_n) \, V(K_2,K_2,K_3\ldots,K_n).
\label{eq:alexandrov-fenchel}
\end{equation}
The Minkowski  functionals of a  body  $K$  in  $d$ dimensions can  be
related to mixed volumes of two bodies  via\footnote{The volume of the
$j$--dimensional unit   ball is denoted  by  $\omega_j$.  Some special
values      are  $\omega_0=1$,    $\omega_1=2$,   $\omega_2=\pi$,  and
$\omega_3=\tfrac{4\pi}{3}$.  The general formula reads
\begin{equation*}
\omega_j=\frac{\pi^{j/2}}{\Gamma\left(j/2+1\right)}.
\end{equation*}
}
\begin{equation}
V_j(K) = \frac{1}{\omega_j} V(
\underset{d-j}{\underbrace{K,\ldots,K}},
\underset{j}{\underbrace{B,\ldots,B}}),
\end{equation}
where $B$ denotes the $d$--dimensional unit ball.  Consult the book by
{\scite{burago:geometric}}   for    a   broad   discussion    of   the
Alexandrov--Fenchel inequality and related issues.

In    three    dimensions,    the    Alexandrov--Fenchel    inequality
(\ref{eq:alexandrov-fenchel})  leads to  two  independent non--trivial
inequalities for the Minkowski functionals, namely
\begin{eqnarray}
\label{eq:isoperimetric}
{V_1}^2 &\geq \frac{\pi}{4} V_0 V_2 , \\
{V_2}^2 &\geq \frac{8}{3\pi} V_1 V_3 . \\
\end{eqnarray}

These  two  inequalities  motivate  the introduction  of  the Blaschke
diagram   {\cite{hadwiger:altes}}.   A  convex   body  with  Minkowski
functionals $V_j$ is mapped to a point $(x,y)$ with coordinates
\begin{equation}
x:=\frac{{\pi}V_0V_2}{4V_1^2},\qquad
y:=\frac{8V_1V_3}{3{\pi}V_2^2}.
\end{equation}
A ball has $x=y=1$.  For  convex bodies, all Minkowski functionals are
non--negative, and  satisfy the inequalities (\ref{eq:isoperimetric}),
so these  points are confined  to the unit square.   Nevertheless, the
convex  bodies do  not  fill the  whole  unit square  -- the  limiting
isoperimetric inequality has  yet to be found (\pcite{schneider:brunn}
and references therein).

The     dimensionless    shapefinders    introduced     in    equation
(\ref{eq:shapefinders}) are related to the isoperimetric ratios via
\begin{equation}
\cP=\frac{1-x}{1+x},\qquad \cF=\frac{1-y}{1+y}.
\end{equation}
Obviously, a  scatter plot of  shapefinders is almost  equivalent to
the ordinary  Blaschke diagram reflected at the  point $(0.5,0.5)$, so
either  method of  presentation should  convey the  same morphological
information.   Throughout   this  article,   we  refer  to   plots  of
shapefinders $(\cP,\cF)$ as Blaschke diagrams.

\providecommand{\bysame}{\leavevmode\hbox to3em{\hrulefill}\thinspace}

%\newpage\section*{Figure captions}

%\newpage\section*{Tables}

\begin{table}[h]
\begin{center}
\begin{tabular}{clcl}
\hline
 & geometric quantity & $\mu$ & $V_\mu$ \\
\hline
$V$    & volume               & 0 & $V_0=V$      \\
$A$    & surface              & 1 & $V_1=A/6$    \\
$H$    & mean curvature       & 2 & $V_2=H/3\pi$ \\
$\chi$ & Euler characteristic & 3 & $V_3=\chi$   \\
\hline
\end{tabular}
\end{center}
\caption{
\label{tab:minkowski}
Minkowski functionals expressed in terms
of the corresponding geometric quantities.}
\end{table}

\end{document}